\documentclass[preprint,showpacs,aps]{revtex4}
\usepackage{graphicx}
\usepackage{bm}
\begin{document}

\title{Collapse transition in mixtures of Bosons and Fermions}

\author{S. T. Chui$^a$}%
\author{V. N. Ryzhov$^b$}

\affiliation{%
$^a$Bartol Research Institute, University of Delaware, Newark, DE 19716, USA\\
$^b$Institute for High Pressure Physics, Russian Academy of
Sciences 142190 Troitsk, Moscow Region, Russia
}

\date{\today}

\begin{abstract}
For mixtures of Bose and Fermi alkali atoms, the Fermion degrees 
of freedom can be integrated out in their {\bf finite temperature}
partition function. The final result can be expanded as a power
series in the Boson density.
Under appropriate conditions,
the pairwise
interaction between the bosons can be changed from positive to
negative at a low enough temperature by the Fermion-mediated term and 
the Boson cloud may then collapse. For attractive Fermion-Boson interactions,
there is also a collapse of the Fermion cloud.
This transition is first order (discontinuous) because the leading order
nonlinear interaction term between the Bosons
is {\bf third order in Boson density with a negative
coefficient}.
We discuss the finite temperature phase diagram of this transition.
Our result may provide for an explanation of
recent experimental observations by Modungo and coworkers.
\end{abstract}


\maketitle
There is much recent interest in mixtures of Bose and Fermi alkali atoms
in traps experimentally.\cite{mo,fs,Had,Roa,mod} This includes different
hyperfine states of the Li6-Li7 system, Na23-Li6 and K40-Rb87. The original
motivation was to study possible superconductivity in the Fermions.
In these systems, the Fermions are essentially non-interacting. From
the Fermi exclusion principle, no two Fermions can be at the same place,
the s-wave scattering between them cannot occur. As a result,
evaporative cooling of the Fermions ${\bf alone}$ is ineffective.
To evaporatively cooled
the Fermions the
Bosons were used as a refrigerant experimentally.
Much of the current theoretical interest is focused on estimating the
possible superconducting properties of the Fermion system.
Nevertheless,
mixtures of Bosons and Fermions are interesting in their own right.
Modugno and 
coworkers\cite{mod} recently studied mixtures of $^{40}K$ and $^{87}Rb$ with 
an attractive interaction between the Boson and Fermion and found 
that as the number of Boson is increased there is a ``instability value''
$N_{Bc}$ at which a 
{\bf discontinuous} leakage of the Boson and Fermion occurs.
They observed a {\bf first
order} collapse transition of the Boson and Fermion clouds.
This paper explores the origin and the nature of this collapse transition.
The nature of the transition depends on the
higher order term of the Boson interaction.
Because there is no direct interaction between the Fermions, it is possible
to formally exactly integrate out their degree of freedom 
in their {\bf finite temperature}
partition function. The final result can be expanded as a power
series in the Boson density.
An effective interaction between the Bosons is obtained.
We found that the pairwise effective interaction can become
attractive for a large enough number of Bosons and at a sufficiently
low temperature and the Boson cloud collapses. 
This is analogous to earlier experiments
with {\bf fixed}
attractive Boson interaction\cite{Rice} where the Boson cloud also
collapses.
In the present case, the effective interaction depends on the 
atom densities and can change during the collapse.
In addition, in contrast to the one component case, the
Boson interaction involve powers of the Boson density higher than the second.
The leading nonlinear interaction term is
proportional to an {\bf odd} (third)
power of the Boson density ($n_B^3$) with a negative coefficient.
As a result the transition is discontinuous
and physical quantities change {\bf discontinuously}. 
Using the experimental parameters, we estimate the instability
Boson number $N_{Bc}$ for
the Boson cloud to collapse. We found $N_{Bc}$ to be 
{\bf extremely sensitive} to the precise value of the Boson Fermion
scattering length $a_{BF}$. For $N_f=2\times 10^4$, $N_{Bc}\approx 10^5$ when
$a_{BF}\approx  -16.7 nm$, in good agreement with 
experimental results.

The Boson-Fermion system may also exhibit a superconducting transition.
The Boson induced Fermi interaction is attractive and thus the
superconducting transition always occurs at zero temperature.
The superconducting transition temperature for the Fermions
is usually {\bf very low}, because it is an ${\bf exponential}$ function
of the dimensionless coupling constant $a$: $T_{sc}=T_F
\exp(-1/a)$. As a result the superconducting transition is difficult
to observe experimentally. Thus it is important to know if the same
difficulty applies to the collapse transition.
The renormalized Boson interaction  can change from attractive
to repulsive at a sufficiently high temperature.
A phase diagram of $N_{Bc}$ as a function
of the temperature T is worked out.
We found that the collapse transition is much easier to observe than the
superconducting transition.
A possible interpretation of the observation of Modugno
and coworkers from
the above picture is as follows:
As a result of the Fermion-Boson attraction, the collapse 
transition will happen to both the Boson and the Fermion.
When the density is high enough, they disappear from the trap.
If the amount of Fermion is smaller than the Bosons, they will be gone first.
Once the Fermion density is small enough, the Bosons will stop collapsing.
We now discuss our results in detail.

We first integrate out the Fermion degrees of freedom in the path integral
for the system.
Our starting point is the functional-integral representation of the
grand-canonical partition function of the Bose-Fermi mixture. It has the form
\cite{popov1,popov2,stoof1,stoof2}:
\begin{equation}
Z=\int D[\phi^*]D[\phi]D[\psi^*]D[\psi]\exp\left\{-\frac{1}{\hbar}\left(
S_B(\phi^*,\phi)+S_F(\psi^*,\psi)+S_{int}(\phi^*,\phi,\psi^*,\psi)\right)\right\}.
\label{1}
\end{equation}
$\phi(\tau,{\bf r})$ describes the Bose component of the mixture, whereas
$\psi(\tau,{\bf r})$ corresponds to the Fermi component.
This integral consists of an integration 
over a complex field $\phi(\tau,{\bf r})$, which 
is periodic on the imaginary-time interval $[0,\hbar\beta]$, and over the Grassmann field
$\psi(\tau,{\bf r})$, which is antiperiodic on this interval. 
The term describing the Bose-gas
has the form:
$S_B(\phi^*,\phi)=\int_0^{\hbar\beta}d\tau\int d{\bf r}\left\{\phi^*(\tau,{\bf r})\left(\hbar
\frac{\partial}{\partial \tau}-\frac{\hbar^2\nabla^2}{2 m_B} +V_B({\bf r})
-\mu_B\right)\phi(\tau,{\bf r})+
\frac{g_B}{2}|\phi(\tau,{\bf r})|^4\right\}.
\label{2}
$
Because the Pauli principle forbids $s$-wave scattering between fermionic atoms in the 
same hyperfine state, the Fermi-gas term can be written in the 
noninteracting form:
$S_F(\psi^*,\psi)=\int_0^{\hbar\beta}d\tau\int d{\bf r}\left\{\psi^*(\tau,{\bf r})\left(\hbar
\frac{\partial}{\partial \tau}-\frac{\hbar^2\nabla^2}{2 m_F} +V_F({\bf r})
-\mu_F\right)\psi(\tau,{\bf r})
\right\}.
$
The term describing the interaction between the two components of the Fermi-Bose
mixture is:
$S_{int}(\phi^*,\phi,\psi^*,\psi)=g_{BF}\int_0^{\hbar\beta}d\tau\int d{\bf r}
|\psi(\tau,{\bf r})|^2|\phi(\tau,{\bf r})|^2,
$
where $g_B=4\pi \hbar^2a_B/m_B$ and $g_{BF}=2\pi \hbar^2a_{BF}/m_I$, $m_I=m_B m_F/(m_B+m_F)$.

The integral over the Fermi fields $\psi$ is Gaussian. 
We can calculate this integral and obtain the partition 
function of the Fermi system as a functional of 
the Bose field $\phi(\tau, {\bf r})$. 
To proceed we
rewrite the partition function as
$
Z=\int D[\phi^*]D[\phi]Z_F\exp\left\{-\frac{1}{\hbar}\left(
S_B(\phi^*,\phi) \right) \right\}.
$
$Z_F$ is the integral involving $\psi(\tau,{\bf r})$:
$Z_F=\int D[\psi^*]D[\psi]\exp\left(-\frac{1}{\hbar}\left(S_F(\psi^*,\psi)+
S_{int}(\phi^*,\phi,\psi^*,\psi)\right)\right)=
\int D[\psi^*]D[\psi]\exp\left\{\int_0^{\hbar\beta}d\tau\int d{\bf r} 
\int_0^{\hbar\beta}d\tau'\int d{\bf r'}\psi^*(\tau,{\bf r})
{\bf G^{-1}}(\tau,{\bf r},\tau',{\bf r'})\psi(\tau',{\bf r'})\right\},
$
where
${\bf G^{-1}}={\bf G^{-1}_F}-{\bf \Sigma}
$
${\bf G^{-1}_F}(\tau,{\bf r},\tau',{\bf r'})=-\frac{1}{\hbar}
\left(\hbar\frac{\partial}{\partial\tau}
-\frac{\hbar^2\nabla^2}{2m_F}+V_F({\bf r})-\mu_F\right)\delta({\bf r-r'})\delta(\tau-\tau');
$
${\bf \Sigma}(\tau,{\bf r},\tau',{\bf r'})=\frac{g_{BF}}{\hbar}|\phi(\tau,{\bf r})|^2
\delta({\bf r-r'})\delta(\tau-\tau').
$

Using the formula for the gaussian integral over the Grassmann variables 
\cite{popov1,popov2,stoof1}
$\int\prod_n d\psi^*_n d\psi_n\exp\left\{-\sum_{n,n'}\psi^*_nA_{n,n'}\psi_{n'}\right\}
=\det A=e^{Tr[\ln A]}
$
$Z_F$ becomes only a function of the Bose fields $\phi$:
$Z_F=\exp\left(Tr\ln\left(-{\bf G^{-1}}\right)\right)
$
${\bf G}$ contains the additional interaction between the Bosons due to the
presence of the Fermions. The properties of the Fermions can also be calculated.
For example, the Fermion density can be calculated from the functional
derivative
$<\psi^*(r)\psi(r)>=<\delta Z_F/\delta G^{-1}(\tau,r,\tau,r)>$.
To illustrate the applicability of this result,
we consider its contribution to the effective interaction between the Bosons.

In the absence of the Fermions, the Boson wave function can be represented
as a wave function $\phi_0$ which is given in the Thomas-Fermi approximation
by $|\phi_0|^2=(\mu_B-0.5m_B\omega_B^2(x^2+y^2+\lambda^2z^2))/g_B$.
Multiboson interaction can be obtained from an expansion of $ln(-{\bf G}^{-1})$
in powers of the deviation of the boson density from $|\phi_0|^2$:
$|\phi|^2=\delta|\phi|^2+|\phi_0|^2$. In this spirit, we write
${\bf \Sigma}={\bf \Sigma_0}+{\bf \delta \Sigma}$ where
${\bf \Sigma_0}(\tau,{\bf r},\tau',{\bf r'})=g_{BF}|\phi_0(\tau,{\bf r})|^2
\delta({\bf r-r'})\delta(\tau-\tau')/\hbar
$.
The corresponding Green's function is defined as
${\bf G_0^{-1}}={\bf G^{-1}_F}-{\bf \Sigma_0}
$.
We have
${\bf G^{-1}}={\bf G^{-1}_0}-{\bf \delta\Sigma}={\bf G^{-1}_0
(I-G_0\delta\Sigma)}$,
$Tr(\ln(-{\bf G^{-1}}))=Tr(\ln(-{\bf G^{-1}_0}))-\sum^\infty_{n=1} Tr[
({\bf G_0\delta \Sigma})^n]/n=-\sum^\infty_{n=0}S_n/\hbar
$.
The pair interaction comes from the term
$S_2=Tr({\bf G_0\delta \Sigma})^2/2$. The resulting expression can be
evaluated in the Thomas-Fermi approximation and assuming that the Fermions
move faster than the Bose condensates. 
To illustrate our approach we consider
the term $S_1= Tr({\bf G_0\delta \Sigma})$. 
Let us rewrite ${\bf G_0}$ in the form:
${\bf G_0}(\tau,{\bf r},\tau',{\bf r}')=\sum_{\omega,n}-\hbar\xi_n({\bf r})\xi_n^*({\bf r}')e^{-i\omega(\tau-\tau')}/(-i \hbar \omega+\epsilon_n
-\mu_F) \hbar\beta
$,
where $\omega=\pi(2s+1)/\hbar\beta$; $s=0,\pm 1,...$,
$\left(-\hbar^2\nabla^2/2m_F+V_F({\bf r})+g_{BF}|\phi_0({\bf r})|^2
 \right)
\xi_n({\bf r})=\epsilon_n\xi_n({\bf r})
$.
$S_1$ becomes
$S_1=Tr {\bf G_0\Sigma}=\int d\tau d{\bf r}{\bf G_0}(\tau,{\bf r},\tau,{\bf r})
g_{BF}\delta |\phi(\tau,{\bf r})|^2/\hbar $
$=g_{BH}/(\hbar\beta)\int_0^{\hbar\beta}d\tau d{\bf r}\delta
|\phi(\tau,{\bf r})|^2\sum_{\omega, n}
\xi_n({\bf r})\xi_n^*({\bf r})/[i \hbar\omega-(\epsilon_n-\mu_F)]
$
$ =g_{BH}/(\hbar\beta)\int_0^{\hbar\beta}d\tau d{\bf r}\delta
|\phi(\tau,{\bf r})|^2\sum_n
\xi_n({\bf r})\xi_n^*({\bf r})\beta/[e^{\beta(\epsilon_n-\mu_F)}+1] 
$.
In the semiclassical Thomas-Fermi approximation \cite{butts} one has:
\begin{equation}
\sum_n
\xi_n({\bf r})\xi_n^*({\bf r})\frac{1}{e^{\beta(\epsilon_n-\mu_F)}+1}=
\frac{1}{(2\pi \hbar)^3}\int d{\bf p}\frac{1}{e^{\beta(H_0({\bf p,r})-\mu_F)}+1},
\label{15}
\end{equation}
where $H_0({\bf p,r})=p^2/(2m_F)+V_F({\bf r})+g_{BF}|\phi_0({\bf r})|^2$. 
We get
\begin{eqnarray}
S_1=
\frac{g_{BF}}{\hbar\beta}\int_0^{\hbar\beta}d\tau d{\bf r}\delta
|\phi(\tau,{\bf r})|^2
\frac{1}{(2\pi \hbar)^D}\int d{\bf p}\frac{1}{e^{\beta(H_0({\bf p,r})-\mu_F)}+1}
\end{eqnarray}

Similarly we get
\begin{eqnarray}
S_2 =-\frac{\beta g_{BF}^2}{2\hbar}\int_0^{\hbar\beta}d\tau d{\bf r}(\delta|\phi(\tau,{\bf r})|^2)^2
\frac{1}{(2\pi\hbar)^D}\int d{\bf p}\frac{e^{\beta(H_0({\bf p,r})-\mu_F)}}{(1+
e^{\beta(H_0({\bf p,r})-\mu_F)})^2}.
\label{17}
\end{eqnarray}
where $H_0({\bf p,r})=p^2/(2m_F)+V_F({\bf r})-g_{BF}|\phi_0|^2$.

We can write the effective Bosonic Hamiltonian in 
a power series in $\delta |\phi|^2$:
\begin{equation}
H_{GP}=\int d{\bf r}\left\{\frac{\hbar^2}{2m_B}|\nabla\phi|^2+V_{eff}({\bf r})
\delta |\phi|^2
+\frac{g_{eff}}{2}(\delta |\phi|^2)^2
+c g_{BF}^3(\delta |\phi|^2)^3\right\},
\label{22}
\end{equation}
where
\begin{eqnarray}
V_{eff}=0.5D \gamma_D g_{BF}\int_0^\infty\frac{\epsilon^{D/2-1}d\epsilon}
{1+e^{\beta(\epsilon-{\tilde \mu})}},
\label{23}\\
g_{eff}=g_{B}-\beta\kappa_D g_{BF}^2\int_0^\infty\frac{\epsilon^{D/2-1}
e^{\beta(\epsilon+V_F-{\tilde \mu})}d\epsilon}
{(1+e^{\beta(\epsilon-\tilde {\mu}})^2}.
\label{24}
\end{eqnarray}
$\tilde{\mu}=\mu_F-V_F({\bf r})-g_{BF}|\phi_0({\bf r})|^2,$
$\kappa_D=(2m_F/\hbar^2)^{D/2}/(2\pi C_D)$, $C_1=1$, $C_2=2$, $C_3=2\pi$;
$\gamma_D=(2m_F/\hbar^2)^{D/2}/(2\pi C'_D)$, $C'_1=1/2$, $C'_2=2$, $C'_3=3\pi$.
$c=(2m_F/\hbar^2)^{D/2}/(12\mu_F^{0.5}\pi C_D)$.
For D=3, we find using the low temperature expansion
\begin{eqnarray}
V_{eff}&=&g_{BF}\left(\gamma_3{\tilde \mu}^{3/2}+\frac{\pi^2}{12\tilde \mu^{1/2}}
\kappa_3
(k_BT)^2\right), \label{28}\\
g_{eff}(T)&=&g_B-g_{BF}^2\kappa_3{\tilde \mu}^{1/2}
\alpha.
\label{37}
\end{eqnarray}
$\alpha=[1-\frac{\pi^2}{24}(k_BT/{\tilde \mu})^2].$

In the same approximation,
we get, for D=3,
$n_F({\bf r})= \kappa_3\int_0^\infty\frac{\sqrt{\epsilon}d\epsilon}
{1+e^{\beta (\epsilon-{\tilde \mu})}}.
$
\%end{equation}
and 
at low temperatures we have:
\begin{equation}
n_F({\bf r})=2\kappa_3\tilde{\mu}^{3/2}/3
+\frac{\pi^2\kappa_3}{12\tilde{\mu}^{1/2}}(k_BT)^2.
\label{32}
\end{equation}
We thus expect the Fermion density $n_F$ to increase as the Boson density 
increases.
As usual, $\mu_F$ can be determined from the equation
$N_F=\int d{\bf r}n_F({\bf r}). 
$


Modugno and 
coworkers\cite{mod} recently studied mixtures of $^{40}K$ and $^{87}Rb$ with 
an attractive interaction between the Boson and Fermion and found a collapse
of the Fermion cloud.  The parameters of 
the system are the following: $a_B=50 \AA$, $ a_{BF}=-217^{+48}_{-43} \AA $,
the aspect ratio of the trapping potential, $\lambda=0.076$.
The corresponding Fermi wave vector is $k_F=7.7\times 10^{-4} /\AA$, 
and $ T_F\approx 360 nK$
when the number of Fermions, $N_f\approx 2\times 10^4$.

Eq. (5) looks like a typical Landau free energy expansion of a system
with order parameter $O=\delta |\phi|^2$ in a weak external field
$V_{eff}$. As we learn from the study of critical phenomena, an instability
occurs when coefficient of the $O^2$ term becomes zero.
This {\bf instability} condition
of $g_{eff}=0$ imposes a condition on ${\tilde \mu}$. Now
$g_{eff}(r=0)=g_B-g_{BF}^2\kappa^3(\mu_F -g_{BF}\mu_B/g_B)^{1/2}$\cite{detail}.
With the Thomas-Fermi approximation
$\mu_B=(m_B\omega_B^2)^{3/5}(15 N_Bg_{B}/4\pi \lambda )^{2/5}/2$.
Substituting this expression of $\mu_B$ into the equation of
$g_{eff}(r=0)=0$, we obtain an estimate of
of the critical Boson number $N_{Bc}$ for the instability
of the system:
\begin{equation}
N_{Bc}=\left( 8\pi[(g_B\pi m_I/g_{BF} m_Fa_{BF}k_F\alpha)^2-1]\mu_Fd_B^{3}
(a_{B}/d_B)^{3/5}/g_{BF}\right)^{5/2} (\lambda /15 ) .
\end{equation}
$d_B=\sqrt{\hbar/m_b\omega _b}$ is the transverse harmonic length 
for the trapping potential of the Bosons.
$\alpha$ is a finite temperature correction factor defined after eq. (9).
The dependence of $N_{Bc}$ on $a_{BF}$ is extremely sensitive.
In Fig. 1 we show $N_{Bc}$ as a function of $a_{BF}$ 
for $T/T_F=0.3$ in units of nm.
$N_{Bc}$ changes by a factor of two when $a_{BF}$
changes by 5 per cent. The experimental value of $N_{Bc}\approx 10^5$
suggests that $a_{BF}\approx -16.7 nm$.

Even when $g_{eff}<0$, there is still a tunnelling barrier for the collapse
to occur. This tunnelling barrier diappears when $g_{eff}=-g_c$ where
estimates of $g_cN_c$ has been given previously as a function
of $\lambda$.\cite{Ueda} We have solved this equation iteratively, 
starting with the initial guess in eq. (28). We found that
the corrections to $N_c$ from this condition is less than 1 per cent.

We next address the nature of the phase transition.
In previous studies of attractive one component Bosons, the free energy
is a function of only the first and the second power of the Boson density.
There are {\bf no} higher power terms.
For the present two component system, the effective Boson interaction
involve higher power terms of the Boson density. The nature of the transition
is determined by these nonlinear terms.
We found a term of the order of $cg_{BF}^3(\delta n_B)^3$ in the
free energy functional.
From Landau type arguments, if an odd order term with a negative coefficient
appears, then at the transition
the free energy can be further lowered by a {\bf finite}
increase of $\delta n_B$; the transition is first order. This is in agreement
with the experimental observation of a discontinuous leakage in the number
of Bosons and Fermions. 
The usual second order phase transition require a leading nonlinear correction
of the order of $n_B^4$ with a positive coefficient and no third order term.
This is not the case in the present system.

The experiment was carried out at finite temperatures.
Upon increasing the temperature the induced
interaction between the Bose particles can change from attractive
to repulsive.  From the condition that $g_{eff}(T)=0$, we obtain
a phase diagram of $N_{Bc}$ as a function of $T/T_F$. This
is shown in Fig. 2.
The current system may also exhibit a superconducting transition.
The Boson induced Fermi interaction is attractive at zero temperature.
The superconducting transition temperature for the Fermions
is usually very low, because it is an ${\bf exponential}$ function
of the dimensional coupling constant $a$: $T_{sc}=T_F
\exp(-1/a)$. As a result the superconducting transition is difficult
to observe experimentally.
The collapsing transition temperature is much higher and do not have
this problem.

Roth\cite{Roth} has studied the structure of the mixture
numerically at zero temperature with a modified Gross-Pitaevskii equation
for the bosons which self-consistently includes the mean-field interaction 
generated by the fermionic cloud. Numerical results were presented for 
the instability Boson number $N_{Bc}$
as a function of $a_{BF}/d_B$ 
for different values of $a_B/d_B$ for {\bf isotropic}
traps. It is {\bf not possible} to compare his
results to the experiment because his results are expressed in terms of a
{\bf single} trap length $d_B$ whereas the experimental trap is
{\bf anisotropic} and there are {\bf two} trap lengths that differ by a factor
of more than three. {\bf When $a_{BF}/d_B$ changes by a factor of three,
$N_{Bc}$ changes by two orders of magnitude.}

Viverit and coworkers\cite{Viverit} have discussed an instability
criterion based on a linear stability analysis at zero temperature.
According to the estimate of ref. (13), the effective
coupling for the experimental system is {\bf repulsive}
and far away from the instability point $g_{eff}/|g_{BF}|=0.055$.
Furthermore their instability criterion does not involve $N_B$.

If the temperature is higher than the Fermi temperature but below
the Bose condensation temperature, it
is still possible to derive an expression for the effective coupling
for the Bosons.
At high temperatures we have the Boltzman limit ($e^{-\tilde{\mu_F}/(k_BT)}\ll1$).
In this case Eqs. (\ref{23}) and (\ref{24}) 
take the form:
\begin{eqnarray}
V_{eff}&=&V_B({\bf r})+\frac{\sqrt{\pi}}{2}g_{BF}
\kappa_3(k_BT)^{3/2}e^{-{\tilde \mu}/(k_BT)}, \label{29}\\
g_{eff}&=&g_B-\frac{\sqrt{\pi}}{2}g_{BF}^2\kappa_3(k_BT)^{1/2}e^{-{\tilde \mu}/(k_BT)}.
\label{30}
\end{eqnarray}
Now the Fermi energy is replaced by the temperature. As expected, 
the renormalization effect becomes very small.

Eqs. (\ref{28}) and (\ref{37}) are useful for studying 
other physical properties of the 
system.  For example, for the
trapped harmonic potential $U_b({\bf r})=\frac 12m_b\omega
_b^2(x^2+y^2+\lambda _b^2z^2)$, the
interaction-induced fractional shift to
the ideal BEC critical temperature 
is \cite{pit} $\Delta T_0/T_0\rightarrow
-1.33(a_{bb}/d_b)(\lambda _bN_b)^{1/6}$
where $N_b$ is the atom number of Bose particles in the trap. 
Replacing the parameters with those obtained
from eq. (18) and (19), our result provides for an easy way to estimate
this change in the presence of Fermions.\cite{ma}

Our result can also be used to calculate the density distribution of the
particles of the mixture.  Using the 
standard Thomas-Fermi approximation with the renormalized interaction
one can obtain the density of the Bose particles. 
The Fermi density can then be obtained from Eq. (\ref{32}) for $T=0$.

$g_{eff}$ only depends on the {\bf square} of the Fermion-Boson
interaction $g_{BF}$. It can become negative for a repulsive $g_{BF}>0$.
In the case, as the Boson density is increased, the Fermion density
{\bf decreases}. 
The system phase separates. After the phase separation is complete
$g_{eff}$ becomes positive again. Thus in that case, we can consider
a negative $g_{eff}$ to be a signature of the instability towards
phase separation. The present calculation provides a unified way of looking
at these two different phenomena.

In this paper, we have used a 3D Thomas-Fermi approximation to describe
the Fermions. This approximation will break down when the anisotropic
quantization
of the energy levels due to the anisotropic 
harmonic potential becomes important.
The larger energy scale corresponds to the transverse
harmonic frequency and is $\hbar\omega_{F,r}=15.16 nK$.
This value is less than the experimental temperature
($T_c=360 nK$). We thus expect the anisotropic
energy quantization effect not to be 
important in the present case. This quantization effect will be important 
in the investigation of possible superconductivity as the temperature
involved is much lower.

In conclusion, we show that because there is no direct interaction
between the Fermions, the part of the path integral involving 
the Fermion field can be carried out exactly. A description
of the Bosons in increasing powers of its density
with renormalized couplings is obtained.
The Fermion-mediated term can rendered
the Boson interaction attractive. When the Fermion-Boson interaction
is attractive, the system collapses. We calculate the
instability Boson number as a function of temperature and 
as a function of the Boson-Fermion scattering length.
The transition temperature is comparable to the Fermi temperature.
Becuase of the structure of the 
leading nonlinear term in the Boson interaction, the transition is first order.
This may provide for an explanation
of recent experimental results of Modugno and coworkers.\cite{mod}

\begin{acknowledgments}
The work was supported in part by the Russian Foundation for Basic
Research (Grant No 02-02-16622). STC is partly supported by a grant from NASA.

\end{acknowledgments}


\begin{thebibliography}{99}
\bibitem{detail} This is true if the radius of the Boson cloud $R_B$
is less than the radius $R_F$ of the Fermion cloud. We have verified
that this is the case.
\end{thebibliography}
\newpage
\begin{figure}
\begin{center}
\includegraphics*[width=8cm,height=6cm]{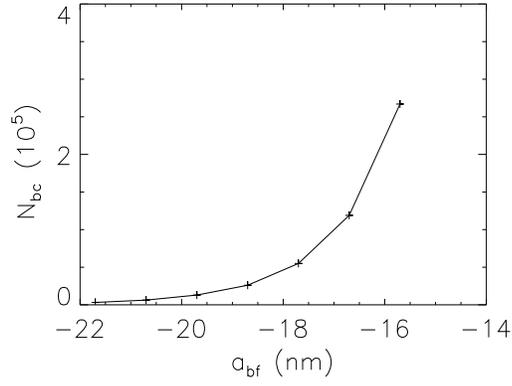}
\caption{The instability Boson number as a function of the Boson-Fermion 
scattering length in units of nm.}
\end{center}
\end{figure}
\begin{figure}
\begin{center}
\includegraphics*[width=8cm,height=6cm]{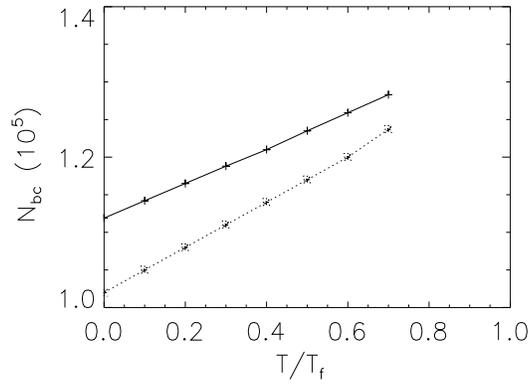}
\caption{The instability Boson number as a function of temperature
normalized by the Fermi temperature for a Boson-Fermion 
scattering length of -16.7 nm for $N_f=2\times 10^4$ (solid line)
and $N_f=5\times 10^4$ (dotted line).}
\end{center}
\end{figure}

\end{document}